# Forking Anatomy: How MorphoDepot Applies the Open-Source Development Model to 3D Digital Morphology

**A. Murat Maga\*, \*\***, Department of Pediatrics, University of Washington, Seattle WA 98195; Center for Developmental Biology and Regenerative Medicine, Seattle Children's Research Institute, Seattle, WA 98101; ORCID: 0000-0002-7921-9018

**Steve Pieper\***, Isomics, Inc., Cambridge, MA 02138; ORCID: 0000-0003-4193-9578

**Cassandra Donatelli**, Department, School of Engineering and Technology, University of Washington, Tacoma, WA 98042; ORCID: 0000-0001-8641-0681

**Paul M. Gignac**; Department of Cellular and Molecular Medicine, University of Arizona College of Medicine, Tucson, AZ, 85724; ORCID: 0000-0001-9181-3258

**Matthew Kolmann**; Department of Biology, University of Louisville, 139 Life Sciences Bldg, Louisville, KY 40292.

**Christopher Noto**, Department of Biological Sciences, University of Wisconsin-Parkside, Kenosha, WI 53144; ORCID: 0000-0003-1829-2593

**Adam Summers**, Department of Biology, University of Washington, Seattle WA 98195

**Natalia Taft**, Department of Biological Sciences, University of Wisconsin-Parkside, Kenosha, WI 53144

\* Joint first authorship
\*\* Corresponding author

## Abstract

The increasing use of 3D imaging technologies in biological sciences is generating vast repositories of anatomical data, yet significant barriers prevent this data from reaching its full potential in educational and collaborative contexts. While sharing raw CT and MRI scans has become routine, distributing value-added segmented datasets—where anatomical structures are precisely labeled and delineated—remains difficult and rare. Current repositories function primarily as static archives, lacking mechanisms for iterative refinement, community-driven curation, standardized orientation protocols, and the controlled terminology essential for downstream computational applications, including artificial intelligence, to help us analyze and interpret these unprecedented data resources.

We introduce MorphoDepot, a framework that adapts the "fork-and-contribute" model—a cornerstone of modern open-source software development—for collaborative management of 3D morphological data. By integrating git version control and GitHub's "social" collaborative infrastructure with 3D Slicer and its SlicerMorph extension, MorphoDepot transforms segmented anatomical datasets from static resources into dynamic, community-curated projects. This approach directly addresses the challenges of distributed collaboration, enforces transparent provenance tracking, and creates high-quality, standardized training data for AI model development. The result is a system that embodies FAIR (Findable, Accessible, Interoperable, and Reusable) data principles while creating powerful new opportunities for remote learning and collaborative science for biological sciences in general and evolutionary morphology in particular.

# Introduction: The Digital Morphology Revolution and Its Unmet Needs

## The Promise of 3D Data

The past two decades have witnessed a profound transformation in how morphologists, comparative anatomists, and specialists in related fields study organismal form [1], [2], [3], [4]. High-resolution computed tomography (CT), magnetic resonance imaging (MRI), and surface scanning technologies have made it possible to non-destructively capture three-dimensional (3D) anatomical structures at resolutions previously unattainable. These techniques have democratized access to rare museum specimens, enabled the study of fragile fossils without physical handling, and opened entirely new avenues for quantitative morphological analysis. As scan resolution and access to equipment grows, we expect an ever-increasing opportunity to apply data driven analysis methods. The revolution in molecular -omics was contemporaneous with, and perhaps even facilitated by, a variety of online repositories that facilitated data-sharing and standardized data quality control procedures. A nascent revolution in phenomics (big anatomical data) is stalled by the lack of such comparable resources, and a limited consensus on what a similar framework could look like for phenotypic data.

## The "Last Mile" Problem: From Data Acquisition to Usability

Despite this technological revolution, a critical gap persists between data acquisition and actual usability. Volumetric scans, stacks of thousands of images representing cross-sectional geometry of the specimen, are merely the starting point. The real scientific and educational value emerges when expert anatomists perform segmentation: the painstaking process of identifying and labeling individual structures voxel by voxel, transforming undifferentiated grayscale volumes into annotated, anatomical atlases that can be queried.

This gap between acquiring raw 3D imaging data and making it scientifically and educationally usable is a "last mile" challenge, i.e., the final, most complex, and expensive component of logistic processes, which has several dimensions. First, segmentation, particularly manual segmentation, is laborious, requiring dozens to hundreds of hours per specimen depending on complexity. Second, it demands deep anatomical expertise—knowing where one structure ends and another begins, or how to trace the boundaries of tissue structures in variable contrast conditions. Often these decisions require knowledge of the physical specimen, including features that may not be directly visible in the scan. Third, and most critically for this discussion, the resulting labeled datasets are trapped in a preservation-versus-innovation gap. As the Community Standards for 3D Data Preservation (CS3DP) framework recognizes, preservation frameworks that excel at long-term archival storage often lag behind the dynamic needs required for active scholarship, iterative refinement, and innovative reuse [5].

Current repositories excel at preserving raw scans and to some extent finalized segmentation in the form of a static 3D model, but they provide no infrastructure for the collaborative improvement of these annotations. Nor is it easy to go back to the segmentation representation from a static 3D model. Errors cannot be easily corrected. Alternative interpretations cannot be proposed and discussed. Datasets cannot evolve as new anatomical insights emerge or terminology standards change. Peer-review is effectively excluded by the framework. In short, the social dynamics that underlie discovery are bogged down by outdated methodology.

The Classroom Challenge: Education in the Age of Digital Specimens

These challenges become even more acute in educational settings. With the increasing availability of free data, instructors may wish to provide students with hands-on experience segmenting real specimens—an invaluable approach that builds both anatomical knowledge and technical skills. However, the logistics of how to acquire dataset files, what software to use, how to collaborate in real-time, and how to track the assignment of who does what can be daunting. The availability of the SlicerMorph ecosystem [6], [7], [8], [9], [10], [11], [12], [13] on the open-source 3D Slicer image computing and analysis platform alleviates majority of the concerns about using a free, robust and open-standards based application, but still the community lacks an open platform for the collaborative creation of complex, multi-layered 3D datasets.

We propose that a fundamental paradigm shift is needed—from conceiving of segmented 3D datasets as static archival products to reconceptualizing them as collaborative, version-controlled projects. The solution lies in adapting a model that has already revolutionized software development: the fork-and-contribute workflow pioneered by Git and popularized by platforms like GitHub.

This paper presents MorphoDepot, the first implementation of this distributed collaboration model for morphological data. By treating each segmentable anatomical dataset as an individual repository that can be forked, modified, reviewed, and merged, MorphoDepot enables communities of researchers, educators, and students to co-create high-quality, standardized anatomical atlases through a transparent, auditable process.

## 2 Background: The Limits of Centralized, Static Archives

The Critical Role of Existing Repositories

Platforms like MorphoSource, Dryad, and institutional repositories have played and continue to play an indispensable role in the digital morphology revolution. They provide persistent identifiers, long-term storage infrastructure, and discovery mechanisms that make 3D data findable and citable. These repositories have successfully moved the field beyond the earlier paradigm by making data readily available and establishing data sharing as a professional expectation and, often, a requirement of grant funding and peer review.

Yet, it is essential to recognize that there is a meaningful distinction between archival preservation—the domain of CS3DP-compliant repositories focused on long-term integrity and accessibility—and working data management—the domain MorphoDepot addresses. These are complementary, not competing, functions. Just as software developers use GitHub for active development but archive major releases on platforms like Zenodo, morphological data benefits from having both collaborative workspaces and permanent archives.

## The Challenge of Creating Usable and Interoperable Labeled Data

**Expertise, Time, and Subjectivity**

Segmentation is not a mechanical process. It requires expert anatomical knowledge to interpret ambiguous boundaries, distinguish between closely apposed structures, and make consistent decisions based on expertise and domain knowledge when tissue contrast is poor. A single cranium might contain dozens of individual bones that may fuse together, each requiring careful delineation across hundreds of CT slices. This may represent many hours of expert labor—a significant investment that should create strong incentives to share and reuse the resulting data.

**Static Datasets and Toolchain Lock-in**
Current repositories typically accept segmented data as "finished" products, which are often 3D model representations derived from the segmentations. These models are often downsampled and smoothened to create visually appealing 3D representation of the specimen in small sample sizes to facilitate easy rendering over the web. Critically, this data is often created using a variety of software platforms, which can be commercial programs with proprietary or undocumented formats. This creates toolchain lock-in: users who download segmented data (assuming it is available) may be unable to modify, correct, or extend it because they lack access to the original software or cannot convert the format to their own program due to lack of documentation. By the framework's very nature, errors become difficult to fix, if not nearly irrevocable.

The situation is exacerbated by lossy format conversions. Because the formats those segmented labelmaps are stored can be proprietary, researchers often resort to converting segmentations to 3D surface models (PLY, OBJ, STL etc.) to improve accessibility. While these formats are widely compatible, they represent a fundamental transformation: the original voxel-based label map—which preserves the full spatial relationship to the underlying grayscale volume it is derived from—is converted into triangular meshes. This conversion can be lossy in different ways. Depending on the pipeline used, fine details might be lost, when the data is decimated or smoothened during the conversion. The ability to easily edit boundaries or extend the segmentation is destroyed. Explicit units of scale of the original data might be lost or may become ambiguous with these formats. Users may have conveniently accessible 3D visualizations but end up losing the data's full analytical potential.

**Lack of Standardized Orientation**
A seemingly simple but profoundly important challenge is anatomical orientation. Research imaging modalities like microCT impose no anatomical meaning on their coordinate systems—a specimen might be scanned upside-down, sideways, or at arbitrary angles depending on how it was positioned in the scanner. Often placement of the specimen into the scanner chamber has more to do with cutting down the imaging time to reduce the costs than anything else. For easier visualisation, comparative analyses, computational atlasing, or machine learning applications, it is much better if data is consistently oriented according to anatomical axes valid for the organism in an explicit way. While at first glance lack of consistent orientation and anatomical system does not seem to impact workflows too much, at the minimum it makes data more difficult to work with and interpret. On the other extreme it may render data useless; if the research question involves naturally asymmetric structures, and the person using it cannot confidently assume what they are seeing on the screen reflects biological reality (e.g., left-right mirroring of the specimen).

**Inconsistent Anatomical Terminology**

Another barrier to data interoperability is inconsistent anatomical nomenclature. Without controlled vocabularies, it is tedious to computationally aggregate or query datasets; or even make sense of what is labeled in the dataset. Comparative anatomy requires manual reconciliation of label names, which can be uninformative if simple ordinal numbers (1, 2, 3, etc.) are used to designate each unique structure, and maybe not even in a consistent manner across datasets. The terms used for labeling anatomical structures need to be associated with specific definitions that can drive the creation of specific segmentation procedures to ensure consistent labeling across specimens.

In MorphoDepot we attempt to address this by requiring that each segmentation label be mapped to standardized anatomical ontologies—particularly UBERON (Uber Anatomy Ontology), when possible [14], [15]. This is enforced during repository creation of "archival" repositories intended for long-term maintenance. We leave up to the data repository owner to determine the specificity of the terminology to be used with their dataset. For example, for a high-resolution scan of a fossil tetrapod skull, the terminology to use might be as simple as "Cranium" (UBERON:0003128), or it might list every cranial bone in a tetrapod skull. We advise the latter approach, which gives them more flexibility should they decide to segment individual cranial bones and fragments in future.

Such consistently labeled annotated data will make developing new algorithms for automated segmentation of multiple species more feasible than it is right now.

## 2.3 The Missing Collaborative Loop

The fundamental limitation of current archival models is their unidirectional nature: data flows from creator to repository to user, but there is no straightforward way to contribute back again. As such the current model is mostly "download and use," but not "download, edit, and contribute back."

This breaks the virtuous cycle that powers improvement in other scientific domains. In ecology and evolution, best practices emphasize data packages that enable validation, extension, and reanalysis. In software, open-source projects thrive because users can report bugs, suggest features, and contribute improvements that benefit the entire community.

Morphological datasets can work the same way. An expert who notices a segmentation error should be able to indicate that, and if they have the time, even propose a correction. A classroom assignment given to students could generate novel data that, after expert review, become part of a permanent reference atlas. Alternative anatomical interpretations can be proposed, discussed, and incorporated through transparent deliberation. None of this is possible when datasets are static files in an archive. What is needed is the infrastructure for collaborative curation—and collaborative curation requires open formats and tools.

## A New Paradigm: The "Fork-and-Contribute" Model

An ideal schema to address this need would employ open, documented file formats that eliminate proprietary file types and dependencies while preserving voxel-based segmentation labels with accurate spatial relationship to source volumes. From a morphological point of view, the schema should enforce anatomical orientations and established ontologies. The infrastructure should support synchronous and asynchronous workflows, allowing independent or group work, alongside transparent provenance tracking as part of the expert review mechanism. The resulting 3D models should be computationally interoperable, ensuring rich opportunities for additional reuse.

The fork-and-contribute model from the domain of open-source software development provides a proven framework, which already meets many of these needs [16]. Its significance in research educational contexts has been long recognized [17], [18], [19], [20], [21]. We have developed MorphoDepot on top of this framework to address the additional, specific needs of CT users and biologists. To understand how MorphoDepot works, it's essential to explain several key concepts from the fork-and-contribute model used in software development

**Repository**: The central project—the "main copy" of the dataset. In MorphoDepot, this contains the source 3D volume, the color table with the custom terminology to be used for segmentation segmented, acquisition and species metadata, and all approved segmentations. The repository lives on GitHub, but is managed through the MorphoDepot interface in 3D Slicer.

**Fork**: An independent copy of the repository created under a collaborator's GitHub account. Contributors don't edit the main copy directly. Instead, they work on their own safe copy. Forks are created automatically when a segmentor (e.g., a student) first loads an assigned task.

**Clone**: A local copy of the forked repository downloaded to the user's computer. This is where the actual work happens—the user loads the 3D volume into Slicer, performs segmentation, and saves their progress locally (i.e., on their home or lab computer).

**Commit and Push**: The act of creating a "save point" that records progress. In MorphoDepot, as you segment, the system automatically tracks which structures you've modified, added, or removed. When you click "Commit and Push," these changes are first committed to the local clone of the repository with a descriptive message about the changes and then this "commit" immediately "pushed" to your fork on GitHub. Commits are what creates the timeline of the repository and the project.

**Pull Request (PR)**: A formal proposal to merge your work back into the main repository. When a student finishes segmenting their assigned structure, they click "Request PR Review." This signals: "My work is ready for inspection." The Pull Request becomes the focus of the review process.

**Review and Merge:** The quality control step. The project owner (instructor, lead researcher) downloads the student's segmentation into Slicer, inspects it in 2D and 3D views, and makes a decision which can be either:

*Approve and Merge:* The segmentation is accurate; it's permanently incorporated into the main repository and the assignment is complete.

*Request Changes:* The work needs improvement and additional changes; feedback is provided, the PR status reverts to "Draft," and the student must revise and resubmit.

This creates a review loop: the student can go to the GitHub repository to see the instructor's feedback, make the necessary changes in MorphoDepot Slicer, commit the changes, and request review again. The loop continues until the work meets standards. Figure 1 summarizes this workflow.

## Core Benefits

The implementation of a "fork-and-contribute" workflow in MorphoDepot facilitates a robust, decentralized environment for collaborative anatomical research with several advantages:

**Decentralized Collaboration:** By utilizing independent forks, contributors can work simultaneously without risking versioning conflicts or compromising the stability of the primary repository. This ensures that the "gold standard" dataset remains protected while experimental or iterative work proceeds in parallel..
**Complete Provenance:** Every change is tracked with attribution, timestamp, and explanation. You can answer questions like "Who segmented the parietal bone?", "When was this boundary modified?", and "Why was this decision made?" This audit trail is invaluable for both research integrity and educational assessment. At the end with a few simple commands, a full history of a MorphoDepot repository can be created as a markdown document.
**Quality Control:** Data integrity is maintained through a mandatory "review-before-merge" protocol. This mechanism ensures that all contributions undergo expert validation (even possibly peer review) before being integrated into the official dataset, effectively preventing the introduction of unvetted or erroneous data.
**Transparent Process:** By hosting technical and interpretive discussions within public version-control threads (e.g., GitHub issues and pull requests), the model creates a permanent, searchable record of the logic underlying specific anatomical interpretations. This transforms the curation process into a transparent scholarly dialogue, providing future researchers with deep context regarding the evolution of the dataset.

# MorphoDepot: Implementation and Workflow

## The Software Ecosystem

MorphoDepot builds on three powerful open-source projects: First, 3D Slicer, aka Slicer, is a medical imaging and analysis platform originally developed for surgical planning and image-guided therapy, now widely used in research across medicine, biology, and paleontology [22], [23]. It provides sophisticated tools for volumetric visualization, segmentation, and measurement—all free and cross-platform (Windows, Mac, Linux). Second, SlicerMorph is an extension to 3D Slicer specifically to handle microCT and related datasets effectively in Slicer [6]. Its extensive tutorial library serves as a knowledge base for organismal biologists and morphologists working with high-resolution datasets like microCTs and other 3D modalities used in research [24]. Third, Git is a free and open-source distributed version control system designed to track changes in files, particularly source code during software development [25] and used by the MorphoDepot as the engine to track the changes.

MorphoDepot itself is implemented as a Slicer extension that adds a new module with tabs corresponding to different phases of the MorphoDepot workflow: Configure, Create, Annotate, Review, and Search. MorphoDepot also uses GitHub and its command line interface (CLI) for the project management aspect of the process [26]. The module handles all Git and GitHub interactions behind the scenes, presenting users with a streamlined interface that requires no command-line knowledge of Git or GitHub tools.

MorphoDepot relies on GitHub for several essential features, such as account management, email notifications, and repository management. Most of these are transparently handled and use of MorphoDepot requires minimal interaction with the GitHub web site. However as a "social coding" site, GitHub offers many useful features that MorphoDepot users may leverage for their work, such as "at mentions" with the @ sign to contact colleagues, "hash tags" with the # sign to reference issues or pull requests, a star icon to "like" a repository, easy to use emojis to quickly express opinions about individual posts, plus a whole range of project management and automation tools provided to address the needs of programming projects. These features, inspired by social media sites, are available, but not required, for MorphoDepot repos.

## MorphoDepot Module Description:

A step-by-step tutorial on using the MorphoDepot module (Figure 2), including an overview of what fork-and-contribute workflow looks like is available at: https://GitHub.com/SlicerMorph/Tutorials/blob/main/MorphoDepot/README.md Here we provide and overview of the module and refer the user to the tutorial linked above for more detailed explanation of the module's full functionality and workflow. Table 1 summarizes the list of actions and where and by whom they need to be taken in the MorphoDepot workflow.

**Configure**

Before creating or contributing to projects, users must configure their system to communicate with GitHub. This involves obtaining a free GitHub account, installing the git command and the

GitHub CLI, gh, to one's computer and authenticating with the GitHub using the CLI. These are one-time steps, and the rest of the MorphoDepot module will not be functional unless all the prerequisite steps are successfully completed. For a quick and easy start with the MorphoDepot, we suggest using the MorphoCloud On Demand Instances [27], [28], a free remote desktop environment with all requisite software and tools are included and the user can get started immediately using the MorphoDepot after completing the GitHub authentication step.

**Create**
This is how a MorphoDepot repository is created on GitHub. There are two mandatory requirements, the 3D scan that will serve as the source volume, and a custom color table with terminology to be used during the segmentation embedded in. Optionally a baseline segmentation can also be included. The rest of the module is a questionnaire that collects metadata about the scan, its contents, and the specimen. The users need to acknowledge that they have the right to redistribute the data (e.g. if they are using a scan that they did not generate themselves) and choose a license under which the others can access their repository, and finally choose a repository name. Screenshots of the scan and 3D renderings of the specimen can also be added to the repository.

Because the source volume included in the repository will define the geometry of all subsequent segmentation and other derivative data, we advise the repository owners to carefully review their scans. If possible, we advise aligning the specimen to match the defined anatomical planes and coordinate system defined in the Slicer (right-left, anterior-posterior, superior-inferior); remove scanning artifacts or packing materials, cropping the extra space while leaving sufficient padding around the boundary of the specimen.
These steps are important since the final volume, when saved on disk in compressed NRRD format, must be 2GB (or smaller), otherwise an error will be generated. Specific steps of how to accomplish these tasks and generating custom color tables using UBERON terminology is explained in the tutorial linked above.

Once a repository is created, the repository owner can add additional information to the standard README.md generated by the MorphoDepot by editing it on the GitHub page.

**Annotate:**
This module is the primary means a segmentor (a student, a remote collaborator etc) will interact with a MorphoDepot repo. To start the process the segmentor needs to open an issue at the repository they want to work on. This triggers a notification to the repository owner, which then assigns the issue back to that individual. After the assignment the segmentor can use the Annotate module to start the task. To prevent users from being spammed with unsolicited issue assignments, GitHub enforces a rule that users must express their interest in contributing to the repo before issues can be assigned. The MorphoDepot module's Annotate tab will show assigned issues and load them on request; actual segmentation is done using the Segment Editor module of the 3D Slicer.

Under the hood, MorphoDepot will fork the original repository under the user's account and where all the changes will take place, until the segmentor is ready to contribute their changes back to the main repository. That's achieved through submitting a Pull Request (PR). The first commit automatically creates a Pull Request in "Draft" status. Subsequent commits update the same PR. When a segmentor is satisfied with their work and ready to contribute back, they change the status from "Draft" to "Ready for Review," by clicking the "Request PR Review"

**Review**
The repository owner receives the PRs ready to be reviewed by clicking "Refresh GitHub" to see all pending PRs. Clicking a PR downloads the proposed segmentation as an overlay on the original volume. The repository owner then uses the general purpose functionality in Slicer to review and visualize the proposed segmentation, while making notes of any changes that are required. They can either request more changes by clicking Request Changes or they can approve the existing segmentation, but clicking Approve Pull Request. This will also merge the segmentation into the main repository.  This loop can repeat as many times as necessary.

The notes section in the Annotate and Review module of the MorphoDepot allows only text entry. However if the discussion would benefit from the graphic depictions, captured screenshots can be added or pasted into the Pull Request thread by going to the GitHub page (the Go to PR Page button).

While there is no formal convention on what the issue titles, issue descriptions or pull request explanation should depict, we suggest careful curation of those both by the segmentor and the repository owner since these notes will form the timeline of how the segmentation is created or modified and become part of the record.

**Search**
The Search tab enables users to discover existing MorphoDepot repositories. After clicking "Load Searchable Repository Data" (which downloads metadata from all public repositories), users can filter the results by keyword, modality, species, sex, and all other metadata entered at the time of repository creation. In the resultant summary table, the user can right-click to either preview the repository in Slicer or go to its GitHub page to see more information.

# Discussion

## Breaking Toolchain Lock-in Through Open-Source Standardization

MorphoDepot's most fundamental contribution may be its demonstration that collaborative curation and file format accessibility are inseparable for digital morphology. By standardizing on 3D Slicer—free, open-source, expandable and cross-platform software—and storing segmentations in NRRD format (a well-documented, open voxel label format), MorphoDepot ensures that anyone, anywhere, with any operating system, can participate in the creation and refinement of datasets. From that point, there are other benefits across multiple domains:

*For Education:* Students can transition from passive consumers of content to active creators of scholarly resources. The skills they learn—version control, collaborative protocols, careful documentation and clear communication—are transferable to modern research and industry contexts. Remote and hybrid learning models become genuinely viable; an instructor in one country can review and provide feedback on student work happening simultaneously across a dozen time zones.

*For Research:* Multi-institution collaborations become dramatically simpler. A paleontologist in Germany and another in Japan can co-segment a holotype specimen without shipping fragile fossils or coordinating file transfers. The complete version history provides transparent provenance—essential for publications and for training researchers who need to understand how anatomical interpretations were reached.

*For Data Quality and Standardization:* The fork-and-contribute model creates powerful incentives for establishing and maintaining standards. Because project owners control what gets merged, they can enforce consistent orientation, terminology, and quality thresholds, at the minimum for their own repos, and their lab projects. Because these standards are documented in the repository (via README and color table files), new contributors can see what's expected. Over time, community norms emerge for shared best practices for segmentation protocols, anatomical definitions, and documentation.

*For Computational Biology and AI:* By enabling collaborative creation of large, version-controlled, community-vetted training datasets with standardized terminology and orientation, MorphoDepot directly addresses the data bottleneck that has prevented AI/ML methods from achieving their potential in morphological analysis. It is not farfetched to imagine MorphoDepot repositories containing hundreds of fish crania from different species, all segmented by a community of students and experts over several years, all using identical UBERON terminology, all in consistent anatomical orientation, all reviewed and approved through the PR process. This is precisely the kind of high-quality, standardized dataset needed to train robust deep learning models for automated segmentation of new specimens. As these datasets proliferate, they could be aggregated to create domain-specific training corpora (mammal skulls, bird wings, primate brains) that accelerate research across entire fields in ways we can currently only imagine and in ways we cannot yet imagine. This is where the impact if MorphoDepot could be most transformative.

## Current Challenges and Limitations

Despite its promise, MorphoDepot faces challenges. For example, MorphoDepot is currently specifically implemented to support the anatomical segmentation problem and does not currently address related research activities such as anatomical landmarking or biomechanical analysis.  However, the open-source nature of the code means that such features could be added either by the original MorphoDepot creators or by the larger community, if there is demand.

For many non-computational biologists, concepts like "fork," "commit," and "pull request" are new. While MorphoDepot abstracts away much of Git's complexity (no command-line operations required for basic workflows), the underlying conceptual model still requires some explanation to the uninitiated. Also, while typical merge conflicts, where two different people edit the same file in different ways, is not possible due to MorphoDepot's design, other types of conflicts can still happen. Resolving them requires understanding the reported error and looking into git log files, which can be challenging for new users. Though, many AI agents can now help resolve them quickly.

GitHub is excellent for active collaboration but is not a certified archival repository. Users still need to take action to mint DOI and upload a static version of their repository to archives like Zenodo to retain a frozen snapshot at time of publication. Though, again, tools exist to do all of those. They are just not yet integrated into MorphoDepot. Additionally, GitHub bulk data attachments have a 2 GB file size limit currently. This is sufficient for many specimens but inadequate for very high-resolution scans. Current workarounds (downsampling, cropping) sacrifice data fidelity. In the future we plan to implement an object store as part of our MorphoCloud project [28], in which the source volumes of MorphoDepot repositories will be stored. The current architecture of the MorphoDepot is flexible to make this transition easily.

Perhaps the biggest criticism that can be made for MorphoDepot is its reliance on a third party, GitHub, to provide the infrastructure. However, it should be noted that this reliance is not unique to MorphoDepot but is shared by almost every other open-source project, including scientific research software projects [17], [19], [21]. While currently all the services MorphoDepot and its repositories rely on are free, this may not be the case going forward. Using GitHub as a front-end was a deliberate choice by the development team so that the limited project resources were spent on delivering technologically complicated workflow in a consistent and easy to use way to the end-users. Otherwise, we would have to provide the entire infrastructure for the project management from scratch, which would be beyond the scope and the resources of our project. It should also be emphasized that while we use GitHub as a platform, particularly for project management, none of the technologies are proprietary or specific to GitHub; they are all open-sourced technologies. Should GitHub change the policy of hosting open-source, public projects for free, it is possible to migrate the entire infrastructure to another platform, or even self-host the project using alternative open-source technologies (e.g., GitLabs) like some other projects do.

## Future Development

Several future development and community engagement could address some of these limitations and expand MorphoDepot's impact and usability such as:

A primary objective may involve the refinement of the user interface to further abstract the complexities inherent in Git-based version control. By further streamlining the integration within 3D Slicer, future updates may seek to minimize the technical burden on researchers through the implementation of automated diagnostics for common issues and the inclusion of native, integrated tutorials. It may make validating that segment names match the color table, that no labels overlap, and that orientation hasn't been inadvertently changed or edit. Such enhancements are designed to facilitate a more intuitive workflow, ensuring that the focus remains on anatomical scholarship rather than infrastructure management.

Release functionality can provide formal connections to archival repositories. For example, MorphoDepot could automatically generate CS3DP-compliant data packages when a project owner creates a "release", then submit these packages to an institutional repositories for permanent archiving with a minted DOI [29], [30].
Current source volume file size limitation can be overcome by supporting large datasets through multi-resolution formats (Zarr, OME-Zarr), cloud-based storage backends, and streaming visualization [31], [32]. This would enable MorphoDepot to handle whole-organism scans of large vertebrates, or extremely large datasets that are coming from emerging fluorescent light-sheet microsopy imaging [33], [34].

Community Standards and sharing involves development of field-specific best practices, e.g., standardized segmentation protocols for particular taxonomic groups, consensus anatomical ontologies, benchmark datasets for validation. We already provide a repository for the community to exchange custom color tables with anatomical terminologies to increase reuse and reduce replicate work [35]. While MorphoDepot provides the infrastructure; ultimately it is the community that must build the consensus and then the standards.

## 6 Conclusion

We present MorphoDepot as a solution to a multifaceted challenge: how to enable distributed, asynchronous collaboration on complex 3D morphological datasets while maintaining quality control, ensuring standardization, preserving provenance, and keeping data accessible in open formats. By adapting the fork-and-contribute model from software engineering, MorphoDepot transforms 3D anatomical datasets from static archived products into dynamic, community-curated projects. The integration of Git version control, GitHub's collaborative infrastructure, and 3D Slicer's open-source segmentation tools creates a complete ecosystem for creating, refining, and sharing high-quality morphological data.

The framework addresses critical gaps in current practice: toolchain lock-in is broken through open-source standardization; orientation chaos is prevented through required data preparation protocols; terminological inconsistency is reduced through enforced ontology mapping; quality control is achieved through structured review workflows. Crucially, MorphoDepot demonstrates that format accessibility and collaborative curation are inseparable. One cannot have genuine community-driven data refinement when the data itself is imprisoned in proprietary formats or lossy conversions. By making open formats and open tools prerequisites for participation, MorphoDepot ensures that every dataset created is not just shared but genuinely reusable.

# LITERATURE CITED

[1] D. M. Boyer, G. F. Gunnell, S. Kaufman, and T. M. McGeary, "MORPHOSOURCE: ARCHIVING AND SHARING 3-D DIGITAL SPECIMEN DATA," *The Paleontological Society Papers*, vol. 22, pp. 157–181, Sept. 2016, doi: 10.1017/scs.2017.13.

[2] T. G. Davies *et al.*, "Open data and digital morphology," *Proc. R. Soc. B*, vol. 284, no. 1852, p. 20170194, Apr. 2017, doi: 10.1098/rspb.2017.0194.

[3] D. C. Blackburn *et al.*, "Increasing the impact of vertebrate scientific collections through 3D imaging: The openVertebrate (oVert) Thematic Collections Network," *BioScience*, vol. 74, no. 3, pp. 169–186, Mar. 2024, doi: 10.1093/biosci/biad120.

[4] P. M. Gignac *et al.*, "The role of networks to overcome large-scale challenges in tomography: The non-clinical tomography users research network," *Tomography of Materials and Structures*, vol. 5, p. 100031, June 2024, doi: 10.1016/j.tmater.2024.100031.

[5] "3D Data Creation to Curation: Community Standards for 3D Data Preservation." Accessed: Dec. 23, 2025. [Online]. Available: https://alastore.ala.org/content/3d-data-creation-curation-community-standards-3d-data-preservation

[6] S. Rolfe *et al.*, "SlicerMorph: An open and extensible platform to retrieve, visualize and analyse 3D morphology," *Methods in Ecology and Evolution*, vol. 12, no. 10, pp. 1816–1825, 2021, doi: 10.1111/2041-210X.13669.

[7] S. Rolfe, C. Davis, and A. M. Maga, "Comparing semi-landmarking approaches for analyzing three-dimensional cranial morphology," *American Journal of Physical Anthropology*, vol. 175, no. 1, pp. 227–237, 2021, doi: 10.1002/ajpa.24214.

[8] A. Porto, S. Rolfe, and A. M. Maga, "ALPACA: A fast and accurate computer vision approach for automated landmarking of three-dimensional biological structures," *Methods in Ecology and Evolution*, vol. 12, no. 11, pp. 2129–2144, 2021, doi: 10.1111/2041-210X.13689.

[9] C. Zhang, A. Porto, S. Rolfe, A. Kocatulum, and A. M. Maga, "Automated landmarking via multiple templates," *PLOS ONE*, vol. 17, no. 12, p. e0278035, Dec. 2022, doi: 10.1371/journal.pone.0278035.

[10] S. M. Rolfe, S. M. Whikehart, and A. M. Maga, "Deep learning enabled multi-organ segmentation of mouse embryos," *Biology Open*, vol. 12, no. 2, p. bio059698, Feb. 2023, doi: 10.1242/bio.059698.

[11] S. M. Rolfe and A. M. Maga, "DeCA: A Dense Correspondence Analysis Toolkit for Shape Analysis," in *Shape in Medical Imaging*, C. Wachinger, B. Paniagua, S. Elhabian, J. Li, and J. Egger, Eds., Cham: Springer Nature Switzerland, 2023, pp. 259–270. doi: 10.1007/978-3-031-46914-5_21.

[12] S. M. Rolfe, D. Mao, and A. M. Maga, "Streamlining asymmetry quantification in fetal mouse imaging: A semi-automated pipeline supported by expert guidance," *Developmental Dynamics*, vol. 254, no. 8, pp. 999–1010, 2025, doi: 10.1002/dvdy.70028.

[13] O. O. Thomas, C. Zhang, and A. M. Maga, "SlicerMorph photogrammetry: An open-source photogrammetry workflow for reconstructing 3D models," *Biol Open*, p. bio.062126, Aug. 2025, doi: 10.1242/bio.062126.

[14] C. J. Mungall, C. Torniai, G. V. Gkoutos, S. E. Lewis, and M. A. Haendel, "Uberon, an integrative multi-species anatomy ontology," *Genome Biol*, vol. 13, no. 1, p. R5, Jan. 2012, doi: 10.1186/gb-2012-13-1-r5.

[15] W. M. Dahdul *et al.*, "A unified anatomy ontology of the vertebrate skeletal system," *PLoS One*, vol. 7, no. 12, p. e51070, 2012, doi: 10.1371/journal.pone.0051070.

[16] M.-W. Wu and Y.-D. Lin, "Open source software development: an overview," *Computer*, vol. 34, no. 6, pp. 33–38, June 2001, doi: 10.1109/2.928619.

[17] A. Zagalsky, J. Feliciano, M.-A. Storey, Y. Zhao, and W. Wang, "The Emergence of GitHub as a Collaborative Platform for Education," in *Proceedings of the 18th ACM Conference on Computer Supported Cooperative Work & Social Computing*, in CSCW '15. New York, NY, USA: Association for Computing Machinery, Feb. 2015, pp. 1906–1917. doi: 10.1145/2675133.2675284.

[18] C. Kamoun, J. Roméjon, H. de Soyres, A. Gallois, E. Girard, and P. Hupé, "biogitflow: development workflow protocols for bioinformatics pipelines with git and GitLab," *F1000Res*, vol. 9, p. 632, 2020, doi: 10.12688/f1000research.24714.3.

[19] E. Escamilla, M. Klein, T. Cooper, V. Rampin, M. C. Weigle, and M. L. Nelson, "The Rise of GitHub in Scholarly Publications," in *Linking Theory and Practice of Digital Libraries*, G. Silvello, O. Corcho, P. Manghi, G. M. Di Nunzio, K. Golub, N. Ferro, and A. Poggi, Eds., Cham: Springer International Publishing, 2022, pp. 187–200. doi: 10.1007/978-3-031-16802-4_15.

[20] Z. Yang, "Git-Based Distributed Collaborative Learning: Theories, Tools and Features," presented at the 2nd International Conference on Education: Current Issues and Digital Technologies (ICECIDT 2022), Atlantis Press, Nov. 2022, pp. 124–133. doi: 10.2991/978-2-494069-02-2_15.

[21] P. H. P. Braga *et al.*, "Not just for programmers: How GitHub can accelerate collaborative and reproducible research in ecology and evolution," *Methods in Ecology and Evolution*, vol. 14, no. 6, pp. 1364–1380, 2023, doi: 10.1111/2041-210X.14108.

[22] A. Fedorov *et al.*, "3D Slicer as an image computing platform for the Quantitative Imaging Network," *Magn Reson Imaging*, vol. 30, no. 9, pp. 1323–1341, Nov. 2012, doi: 10.1016/j.mri.2012.05.001.

[23] R. Kikinis, S. D. Pieper, and K. G. Vosburgh, "3D Slicer: A Platform for Subject-Specific Image Analysis, Visualization, and Clinical Support," in *Intraoperative Imaging and Image-Guided Therapy*, Springer, New York, NY, 2014, pp. 277–289. doi: 10.1007/978-1-4614-7657-3_19.

[24] "Tutorials/README.md at main · SlicerMorph/Tutorials." Accessed: Dec. 23, 2025. [Online]. Available: https://github.com/SlicerMorph/Tutorials/blob/main/README.md

[25] "Git." Accessed: Dec. 23, 2025. [Online]. Available: https://git-scm.com/

[26] "Manual," GitHub CLI. Accessed: Dec. 23, 2025. [Online]. Available: https://cli.github.com/manual/

[27] "MorphoCloud: On-Demand Cloud for 3D Slicer & SlicerMorph," MorphoCloud: On-Demand Cloud for 3D Slicer & SlicerMorph. Accessed: Dec. 20, 2025. [Online]. Available: https://morphocloud.org/

[28] A. M. Maga and J.-C. Fillion-Robin, "MorphoCloud: Democratizing Access to High-Performance Computing for Morphological Data Analysis," Dec. 24, 2025, *arXiv*: arXiv:2512.21408. doi: 10.48550/arXiv.2512.21408.

[29] A. Decan, T. Mens, P. R. Mazrae, and M. Golzadeh, "On the Use of GitHub Actions in Software Development Repositories," in *2022 IEEE International Conference on Software Maintenance and Evolution (ICSME)*, Oct. 2022, pp. 235–245. doi: 10.1109/ICSME55016.2022.00029.

[30] M. Wessel, T. Mens, A. Decan, and P. R. Mazrae, "The GitHub Development Workflow Automation Ecosystems," in *Software Ecosystems: Tooling and Analytics*, T. Mens, C. De Roover, and A. Cleve, Eds., Cham: Springer International Publishing, 2023, pp. 183–214. doi: 10.1007/978-3-031-36060-2_8.

[31] "Zarr," Zarr. Accessed: Dec. 01, 2024. [Online]. Available: https://zarr.dev/

[32] J. Moore *et al.*, "OME-NGFF: a next-generation file format for expanding bioimaging data-access strategies," *Nat Methods*, vol. 18, no. 12, Art. no. 12, Dec. 2021, doi: 10.1038/s41592-021-01326-w.

[33] M. Kumar, J. Nasenbeny, and Y. Kozorovitskiy, "Low cost light-sheet microscopy for whole brain imaging," in *Three-Dimensional and Multidimensional Microscopy: Image Acquisition and Processing XXV*, International Society for Optics and Photonics, Feb. 2018, p. 104991I. doi: 10.1117/12.2288497.
[34] A. K. Glaser *et al.*, "A hybrid open-top light-sheet microscope for versatile multi-scale imaging of cleared tissues," *Nat Methods*, vol. 19, no. 5, Art. no. 5, May 2022, doi: 10.1038/s41592-022-01468-5.
[35] S. Project, *SlicerMorph/terms-and-colors*. (June 09, 2025). Accessed: Dec. 23, 2025. [Online]. Available: https://github.com/SlicerMorph/terms-and-colors

**TABLES**

**Table 1.** A breakdown of what, who, and where for each stage in the MorphoDepot workflow.

| What | Who | Action(s) | Where |
|---|---|---|---|
| Setup | Everyone | Install git and GitHub CLI, then authenticate via gh commands.<br>Install Slicer, its MorphoDepot extension and configure it. | On the computer to be used with MorphoDepot.<br>Configure tab (MorphoDepot) |
| Find repositories | Everyone | Search GitHub for specific MorphoDepot repos using keywords and filters.<br>Preview contents of those repos. | Search tab (MorphoDepot) |
| Creation | Repository Owner | Upload 3D scan, custom color table (UBERON), and specimen metadata | Create tab (MorphoDepot) |
| Create Issue | Contributor | Open a GitHub issue to request a task. | On repository's GitHub page |
| Assignment | Repository Owner | Assign the issue to the issue creator. | On repository's GitHub page |
| Do the work | Contributor | Create a new fork of the repo, get the issue and work on it. Use standard Slicer tools to do the work. | Annotate tab (MorphoDepot)<br>Segment Editor (Slicer) |
| Submission | Contributor | "Commit and Push" changes to push them into user's fork, then submit a Pull Request (PR) | Annotate Tab (MorphoDepot) |
| Evaluation and feedback | Repository Owner | Download proposed segmentation for inspection.<br>Use existing Slicer tools to compare (if applicable).<br>Request changes if necessary | Review Tab (MorphoDepot)<br>Segment comparison (Slicer) |
| Finalization | Repository Owner | Approve and Merge (incorporate into main repo) | Review Tab (MorphoDepot) |

# FIGURES

**Figure 1.** Overview of the MorphoDepot fork-and-contribute workflow. Blue figures represent the Project Maintainer (repository owner), while orange figures represent third-party contributor(s) (e.g., students, collaborators, enthusiasts) that discovered the repository through Search module. The process begins when a contributor creates an issue, which is then assigned back to them by the maintainer. The contributor forks the project to work on the task via the Annotate module, tracking their progress through a series of commits. After submitting a pull request, the owner and contributor collaborate through a review and revision process until the completed work is merged back into the main repository.

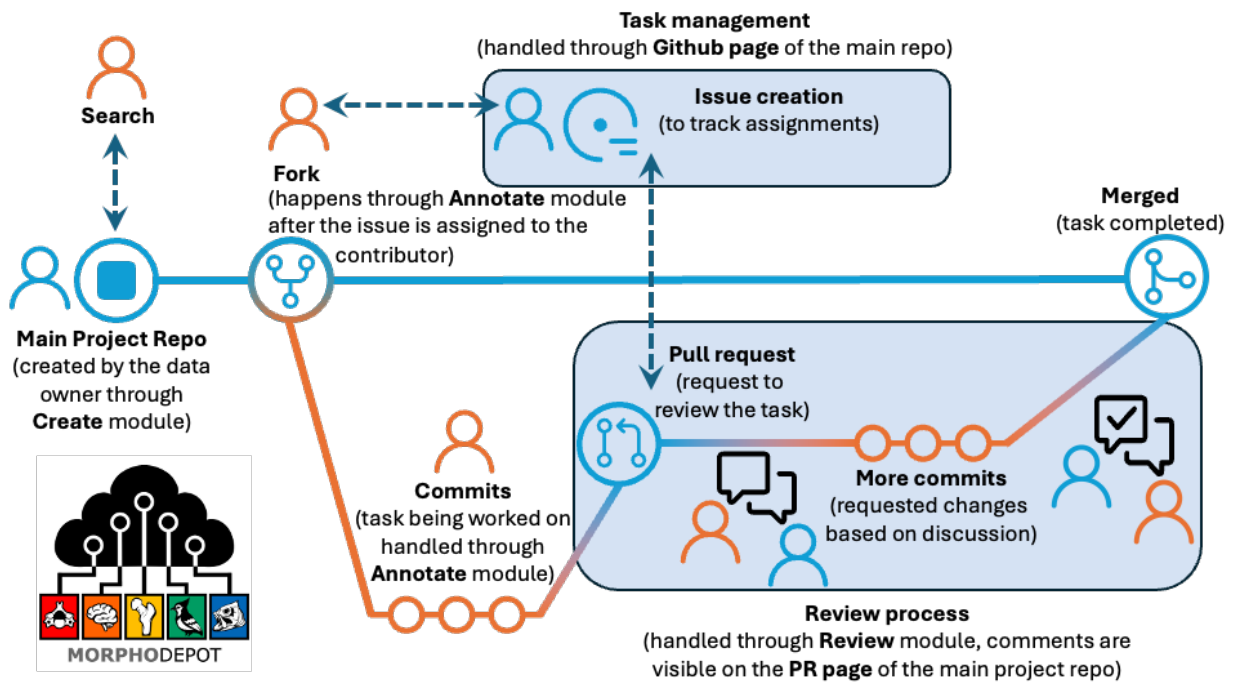

**Figure 2.** Screenshot of the MorphoDepot module as deployed on a MorphoCloud instance. The user is previewing a repository they discovered through the Search functionality.

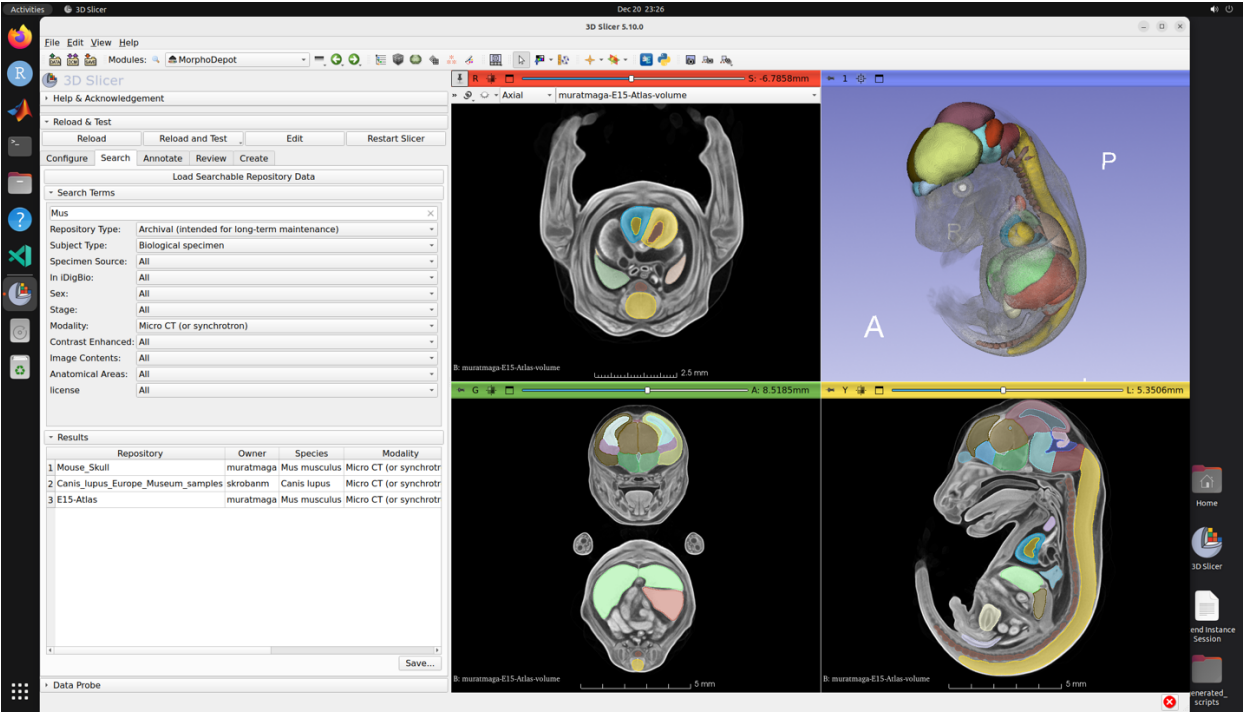